\documentclass[a4paper,11pt]{article}
\usepackage[utf8]{inputenc}

\usepackage{amsfonts}
\usepackage{amssymb}
\usepackage{graphicx}
\usepackage{amsmath}
\usepackage{enumerate}
\usepackage{color}
 \usepackage{multirow}
\usepackage{float}

\RequirePackage{mathrsfs} \RequirePackage[sc]{mathpazo}
\RequirePackage{wasysym} \RequirePackage{setspace}\textheight=650pt
\title{On Hawking radiation of  $3D$ Rotating Hairy Black Holes  }
\author{A. Belhaj$^{1,2}$, M. Chabab$^2$, H. EL Moumni$^2$, K. Masmar$^2$, M.  B. Sedra$^{3}$ \\
\\
{\small $^{1}$D\'epartement de Physique, Facult\'e
Polydisciplinaire, Universit\'e Sultan
 Moulay Slimane, B\'eni Mellal,  Morocco. } \\
{\small $^{2}$High Energy Physics and Astrophysics Laboratory, FSSM,
 \small Cadi Ayyad University, Marrakesh, Morocco.
} \\
{\small $^{3}$  D\'{e}partement de Physique, LHESIR, Facult\'{e} des
Sciences, Universit\'{e} Ibn Tofail,
 K\'{e}nitra, Morocco.} 
 }
\date{\today }
\begin{document}

\maketitle

\abstract{ {

We  study  the Hawking radiation of $3D$  rotating hairy black
holes. More  concretely,     we compute  the transition probability
of a bosonic and fermionic particle in  such  backgrounds. Thew, we
show that the transition probability is independent of the nature of
the particle. It is observed  that  the charge of the scalar hair
$B$ and   the rotation parameter $a$ control such a  probability. }

\newpage


The study of $(1+2)D$-dimensional black holes provides  a good
understanding  of low-dimensional gravity models  and their related
quantum field  theories.  The  significance comes from 
  the evaporation of the black
hole considered as a consequence of Hawking radiation
 \cite{1,2,epl1,edi3,edi4,edi5,edi10,edi12,edi39}.

The  $(1+2)D$ hairy black holes and  their  thermodynamical
properties are  extensively  investigated
  using different methods \cite{Ir, Xu, ZXZ,XZ,Pourhassan:2015jja}. More precisely,
    their statistical physics \cite{our1} and
  critical behavior have been studied by considering the cosmological
   constant as a thermodynamical
  pressure \cite{our2,our, our4,ana1,ana2,ana3}.  It is  interesting  to  recall
   the   connection  between
   such
    $3D$ black holes and non-commutative
   geometry  which  can be found in \cite{ss8,edi38} and with  the CFT correspondence
    as proposed
 in \cite{x7,x9}.\\
 The aim  of this work is to contribute to this program by exploring the
  corresponding Hawking radiation.
 After an introduction of the
rotating hairy $(2+1)D$ black hole, we  discuss  the tunneling of
the bosonic and fermionic particles from such  black holes. Then, we
show that  the probability of the transition is independent of the
the nature of the particle.

To start,  we   reconsider the study of
   $3D$-dimensional gravity with   a
non-minimally coupled scalar field.   The   black hole  solution is
dubbed  hairy black hole in three dimensions. In the absence of
the  Maxwell   gauge fields, the corresponding  model  is 
controlled  by the following action $\cite{ZXZ}$
\begin{equation}
\mathcal{I_R}=\frac{1}{2}\int d^3x
 \sqrt{-g}\left[R -g^{\mu\nu}\nabla_\mu\phi\nabla_\nu\phi-\xi
 R\phi^2-2V(\phi)\right],
\end{equation}
where $\phi$ is  the  dynamical scalar field.  For simplicity
reason, we   elaborate   a particular   situation where  the
coupling and the gravitational constants  are fixed to
$\xi=\frac{1}{8}$ and $\kappa=8\pi G=1$ respectively.   Within this
assumption, the   metric solution takes the following form
\begin{equation}\label{metric}
ds^2=-f(r) dt^2+\frac{1}{f(r)}
dr^2+r^2\left(d\theta+\omega(r)dt\right)^2.
\end{equation}
The  functions $f$ and $w$, describing  such a  black hole solution,
read as
\begin{equation}\label{fr}
f(r)=3\beta+\frac{2B\beta}{r}+\frac{(3r+2B)^2 a^2}{r^4}+\frac{r^2}{\ell^2}\end{equation}
\begin{equation}\label{omega}
\omega(r)=-\frac{(3r+2B)a}{r^3}
\end{equation}
where  $\beta$ is identified with $-\frac{M}{3}$. In these
equations, $a$  denotes  the rotating  angular momentum parameter.
 $B$ is  linked with  the dynamical  scalar field via the relation
\begin{equation}
\phi(r)=\pm\sqrt{\frac{8B}{r+B}}.
\end{equation}
For later use, an explicit form of the potential are  needed. In the
present study, we  use   the potential $V(\phi)$ proposed in
\cite{XZ} and given by
\begin{equation}
V(\phi)=\frac{1}{512} \left(\frac{a^2 \left(\phi ^6-40 \phi ^4+640 \phi ^2-4608\right)
   \phi ^{10}}{B^4 \left(\phi ^2-8\right)^5}+\phi ^6 \left(\frac{\beta
   }{B^2}+\frac{1}{\ell^2}\right)+\frac{1024}{\ell^2}\right).
\end{equation}
Concretely,  we  discuss the Hawking radiation of  the corresponding
black solution.  Indeed, we compute the probability transition of
bosonic and fermerionic particles in such black hole backgrounds.

Here, we  deal with  the emission of scalar particles from rotating
hairy three dimensional black hole as tunneling phenomena across the event horizon.
  This   can be  done by solving the Klein-Gordon equation for the the scalar
  $\Psi$. This equation reads as
  which reads
\begin{equation}\label{KG1}
g^{\mu\nu}\partial_\mu\partial_\nu \Psi-\frac{m^2}{\hbar^2}\Psi=0
\end{equation}
where $\mu$ and $\nu$  take the values $0,1,2$ for the coordinates
$t$,$r$,$\theta$ respectively. The quantity  $m$
 represents  the mass  of the particle.  In  what follows,
  we use $WKB$ approximation and choose an ansatz of the form
\begin{equation}
\Psi(t,r,\phi)=e^{\frac{i}{\hbar}I(t,r,\theta)+I_1(t,r,\theta)+\mathcal{O}(\hbar)}
\end{equation}
Using   (\ref{KG1}),  the  exponential term and multiplying by
$\hbar^2$, we obtain
\begin{equation}
g^{tt}(\partial_t I)^2+ g^{rr}(\partial_r I)^2+g^{t\theta}(\partial_t
 I\partial_\theta I)+g^{\theta\theta}(\partial_\theta I)^2+m^2=0.
\end{equation}
where  $g^{\mu\nu}$ are the components of the metric (\ref{metric}).
Exploring the symmetries of  such  black holes,
  $\partial_t$ and  $\partial_\phi$ are considered  as  the Killing fields.  Thus,  there
exists a solution for this differential equation given  in
terms of the classical action $I$. The later can be written as
follows
\begin{equation}
I=\zeta t + W(r)+ j\theta+K,
\end{equation}
where $w$ and $j$  represent  the energy and angulaire momentum of
the particle respectively. $K$ is a constant which could be complex.
Using the above expressions, we  arrive to
\begin{equation}
W'(r)=\pm\sqrt{-\frac{gtt}{grr}\left(\zeta^2+\frac{g^{\theta\theta}}{g^{tt}}j^2-\frac{g^{t\theta}}{g{tt}}j\zeta+\frac{1}{g^{tt}}m^2\right)}
\end{equation}
Substituting the value of the metric tensor,  we obtain  the
following  integral
\begin{equation}
W_\pm(r)=\pm\int\frac{\sqrt{\zeta^2-(\frac{f(r)}{r^2}-\omega^2)j^2+2\omega\zeta
j-f(r)m^2}}{f(r)}dr.
\end{equation}
We observe a  simple pole at $r=r_+$.  From the residue theory,
dealing with  semi circles,  the performed integration results in,
\begin{equation}\label{h3.8}
W_\pm=\pm \pi i \frac{\sqrt{\zeta^2+(\omega(r_+))^2j^2+2 \omega(r_+)\zeta j}}{f'(r)}.
\end{equation}
Therefore, this  equation implies that
\begin{eqnarray}\label{h3.9}
Im W_+&=&\pi \frac{\zeta+j\omega(r_+))}{f'(r_+)}\\
&=&\frac{3 \pi  r_+^2 \ell ^2 \left(\zeta  r_+^3-2 a j
   \left(B+r_+\right)\right)}{\ell ^2 \left(2 B M r_+^3-3 a^2 \left(8 B+9
   r_+\right)\right)+6 r_+^6}.
\end{eqnarray}
It is recalled that the Hawking radiation  can be viewed as a
process of quantum tunneling of particles from the black hole
horizon. From this point of view,  we  compute  the imaginary part of
the classical action for this classically forbidden process of
emission across the horizon. In this semi-classical approach,  the
probabilities of the crossing the horizon from inside to outside 
and from outside to inside,  reads as  \cite{h7,h8}
\begin{eqnarray}
P_{emission}&\propto& \exp\left(\frac{-2}{\hbar}Im I\right)=\exp\left(\frac{-2}{\hbar}(Im W_+ +Im K)\right).\\
P_{absorption}&\propto& \exp\left(\frac{-2}{\hbar}Im
I\right)=\exp\left(\frac{-2}{\hbar}(Im W_- +Im
K)\right).\end{eqnarray} 

It is known that  any outside particle will
certainly fall into the black hole. Thus,   we should  take $Im K=-
Im W_-$.  Using (\ref{h3.8}), we also have  $W_+=-W_-$. This indicates
that   the probability of a particle tunneling from inside to
outside the horizon is
\begin{equation}\label{h3.12}
\Gamma=\exp\left(\frac{-4}{\hbar} Im W_+\right).
\end{equation}
 A priori there are many ways,  including the  first
thermodynamical  law, to get   the tunneling probability of scalar
field. However we remark, from  (14),   that  $Im W_+$ is
related to the temperature  hidden in the  $f'(r_+)$ expression.
Using Boltzmann factor and substituting (\ref{h3.9})  into (\ref{h3.12}),
we get the desired formula, 
 \begin{equation}\label{prob}
 \Gamma=\exp\left[-\frac{12 \pi  r_+^2 \ell ^2 \left(\zeta  r_+^3-2 a j
   \left(B+r_+\right)\right)}{\hbar  \left(\ell ^2 \left(2 B M r_+^3-3 a^2
   \left(8 B+9 r_+\right)\right)+6 r_+^6\right)}\right].
 \end{equation}
 This quantity  represents  the tunneling probability of scalar particle from inside
  to outside the event horizon of rotating hairy black hole.
 Comparing this equation    with $\Gamma=\exp(-\beta \omega)$,  which
 is Boltzmann factor for particle of energy $\omega$  where  $\beta$ is the inverse
  of the temperature of the horizon \cite{h7,h8}, it is possible to  derive  the Hawking
  temperature. Notice that one can also compute the Hawking temperature directly from the equation,
  \begin{equation}
  T_H=\frac{f'(r_+)}{4\pi}|_{r=r_+}.
  \end{equation}
 and it is easy to cheek that
  \begin{equation}\label{ttH}
  T_H=\frac{3 \left(4 a^2 B^2 \ell ^2+12 a^2 B r \ell ^2+9 a^2 r^2 \ell ^2+r^6\right)}{r^3 \ell
   ^2 (2 B+3 r)}.
  \end{equation}
Besides, similar form to (\ref{ttH}) can be obtained from the entropy using the  first law
of the thermodynamics.

 It is interesting to note here that  we
 can also  recover the formula given in \cite{Ir}.  Furthermore, for  $B=0$  the result  of \cite{tun} is easily reproduced.

Having discussed the bosonic equations, we move now discuss the
fermionic  particles. In this case,  we compute Hawking radiation
 from rotating three dimensional hairy black hole.  Precisely,  we consider
  the two component massive spinor field $\Psi$, with mass $m$,
     verifying  the following  Dirac equation
\begin{equation}\label{dirac}
i\hbar \gamma^\alpha e_a^\mu\nabla_\mu\Psi-m\Psi=0.
\end{equation}
 $\nabla $ is the spinor covariance derivative given by
  $\nabla_\mu=\partial_\mu+\Omega_\mu$, with
 \begin{equation}
 \Omega_\mu=\frac{i}{2} \Gamma_\mu^{\alpha\beta}\Sigma_{\alpha\beta},
 \end{equation}
\begin{equation}
\Sigma_{\alpha\beta}=\frac{i}{4}[\gamma^\alpha,\gamma^\beta],\quad\quad \Omega_\mu=\frac{-1}{8}\Gamma_\mu^{\alpha\beta}[\gamma^\alpha,\gamma^\beta].
\end{equation}
 The $\gamma$ matrices, in three space-time
 dimensions,  take the following form
 \begin{equation}
 \gamma^a=(-i\sigma^1,\sigma^0,\sigma^2)
 \end{equation}
where $\sigma^i$ are the Pauli sigma matrices, and where  $e_a^\mu$
are  the vielbein fields. Choosing  the curved space $\gamma^\mu$
matrices as
\begin{equation}
\gamma^t=\left(\begin{array}{cc}0 & -\frac{1}{\sqrt{f}} \\ \frac{1}{\sqrt{f}} &
0\end{array}\right),\quad \gamma^r=\left(\begin{array}{cc}0 & \sqrt{f} \\\sqrt{f} &
0\end{array}\right),\quad \gamma^\theta=\left(\begin{array}{cc}\frac{1}{r} &
 \frac{8\omega}{\sqrt{f}} \\-\frac{8\omega}{\sqrt{f}} & \frac{1}{r}\end{array}\right)
\end{equation}
satisfying  the condition
$\{\gamma^\mu,\gamma^\nu\}=g^{\mu\nu}\mathbb{1}$, where $\mathbb{1}$
is the identity matrix, the equation of motion   reads as
\begin{equation}\label{6.6}
i\left(\gamma^t\partial_t+
\gamma^r\partial_r+\gamma^\theta\partial_\theta\right)\Psi-\frac{m}{\hbar}\Psi=0.
\end{equation}
For a fermionic particle with  spin $1/2$,    we  have two states
namely spin-up $(\uparrow)$ and spin-down $(\downarrow)$. Thus,  we
can use the following ansatz for the solution
\begin{equation}\label{6.7}
\Psi_\uparrow=\left(\begin{array}{c}X(t,r,\theta) \\ 0\end{array}\right)e^{\frac{i}{\hbar}I_\uparrow(t,r,\theta)}
\end{equation}
\begin{equation}
\Psi_\downarrow=\left(\begin{array}{c}0 \\ Y(t,r,\theta)\end{array}\right)e^{\frac{i}{\hbar}I_\downarrow(t,r,\theta)}
\end{equation}
where $\Psi_\uparrow$ denotes the wave function of the spin-up particle and $\Psi_\downarrow$
 is for the spin-down case. Inserting equation (\ref{6.7}) for the spin-up
 particle into the Dirac equation (\ref{6.6}), then dividing by exponential term
 and multiplying by $\hbar$, we get the following equation
\begin{equation}
-\frac{X}{\sqrt{f}}\partial_\uparrow I_\uparrow(t,r,\theta)
-\sqrt{f}\;\partial_r I_\uparrow(t,r,\theta)+
 \frac{\omega}{\sqrt{f}}\partial_\theta I_\uparrow(t,r,\theta)=0.
\end{equation}
Notice here that the separation of  variables for the spin-up case
can be used to produce  Killing vectors
$\left(\frac{\partial}{\partial t}\right)^\mu$ and
$\left(\frac{\partial}{\partial \theta}\right)^\mu$.  To handle the
above equation, we should  express $I_\uparrow(t,r,\theta)$ as
\begin{equation}\label{6.10}
I_\uparrow(t,r,\theta)=-\zeta t+ W(r)+\Theta(\theta)+K=-\zeta t +W(r) +j \theta +K
\end{equation}
where  $\zeta$ and $j$ are the energy and the angular momentum of
the emitted particle respectively. $K$ is a complex constant. Using
the  expression in the above equation, we obtain
\begin{equation}\label{6.11}
\frac{X}{\sqrt{f}}\zeta -\sqrt{f}\; X \;\partial_r
W-\frac{\omega}{\sqrt{f}}\partial_\theta\Theta=0.
\end{equation}
Putting  the equation (\ref{6.10}) in (\ref{6.11}), we get
\begin{equation}
\frac{\zeta}{\sqrt{f}}-\sqrt{f}\; X \;\partial_r W-j\frac{\omega}{\sqrt{f}}=0
\end{equation}
where
\begin{equation}\label{6.13}
\partial_r W=\frac{1}{f}\left( \zeta -j \omega\right)
\end{equation}
For spin-down particle, the phase $I_\downarrow$ and its
$r$-dependence can be obtained using similar method and steps. More precisely,  we get expressions similar to those shown in  equations (\ref{6.10}) and (\ref{6.13}). Using (\ref{6.13}), we obtain
\begin{equation}
W=\int \frac{d r}{f}(\zeta-j\;\omega).
\end{equation}
Then, we integrate along a semi circle around the pole at $r_+=0$.
At the  horizon of the black hole,  the radial function is then given  by
\begin{equation}
W= \frac{3 \pi  r_+^2 \ell ^2 \left(\zeta  r_+^3-2 a j
   \left(B+r_+\right)\right)}{\ell ^2 \left(2 B M r_+^3-3 a^2 \left(8 B+9
   r_+\right)\right)+6 r_+^6}.
\end{equation}
Similar calculation shows that tunneling probability  takes  the
form   (\ref{h3.12})
\begin{equation}
\Gamma= \exp\left[-\frac{12 \pi  r_+^2 \ell ^2 \left(\zeta  r_+^3-2 a j
   \left(B+r_+\right)\right)}{\hbar  \left(\ell ^2 \left(2 B M r_+^3-3 a^2
   \left(8 B+9 r_+\right)\right)+6 r_+^6\right)}\right].
\end{equation}
Remarkably we find the  same  expression as the one obtained by solving the
Klein-Gordon equation.  It is recalled that similar steps 
to those in the  bosonic case have also been used. Again note that if we set $B=0$,  we recover the result found
in \cite{tun}.

In this work,   we  have studied   the Hawking radiation of $3D$
rotating hairy black holes.  In particular,    we  have computed
the transition probability of  bosonic and fermionic particles in
such backgrounds. Then, we have  shown  that the transition
probability is independent of the nature of the particle. It should
be interesting to extend this work to higher dimensional cases.

\end{document}